# Slip Resistance Test Apparatus of Synthetic Rubber Trackpad on Photovoltaic Surface

Anh Duy Hoang Ngoc, Cong Toai Truong, Minh Tri Nguyen, An Nguyen Danh
Van Tu Duong, Huy Hung Nguyen and Tan Tien Nguyen

*Abstract* – The increasing development of the solar energy industry in many countries has led to a rising frequency of human and robot presence in this area. To ensure occupational safety, various protective equipment, including rubber material, is commonly used for slip resistance while moving on the surface of solar PV panels. Therefore, the slip resistance test apparatus is built for testing the slip resistance between the synthetic rubber trackpad and the photovoltaic panel (PV) surface. Synthetic rubber is a man-made material, so it is difficult to control the parameters of its mechanical and chemical properties absolutely. Variations in wet/dry working conditions or Shore hardness are factors that make slip computation more challenging. Therefore, an apparatus with the principle of converting the reciprocating motion of the screw and the casters into the rotation of the hinge is introduced to adjust the tilt angle of the upper surface, detect and evaluate the slippage of the rubber trackpad by sensors. Some parameters related to accuracy such as vibration and theoretical-empirical assessment, are also mentioned. In addition to designing a reliable apparatus, the article also succeeded in providing a safety standard for synthetic rubber with Shore A30-A40 when moving on PV surfaces.

*Keywords:* slip resistance, slip safety, slip and fall accident, ramp test, pendulum method, photovoltaic, Shore hardness.

## I. INTRODUCTION

The slip resistance refers to the ability to prevent slip, enabling the shoe-floor to move on horizontal or inclined planes in dry, wet, or contaminated working conditions [1]–[7]. Despite the fact that the slip resistance is measured based on the static friction force and the conventional theoretical computation, there are still a lot of factors that affect coefficients of friction (COF). Typical factors that are not mentioned and are not practical to take into account in COF computation include the change in environmental temperature, the roughness of two surfaces in contact, dust accumulation, dry/wet condition, and the viscosity of oil in a polluted environment [5], [8]–[12].

Moreover, most anti-slip shoe-floor are made of elastic materials, specifically rubber, thanks to its characteristics. Many researches also focus on improving slip resistance for rubber[8], [13]–[15], and the Shore hardness is a typical factor that significantly impacts slip resistance. Types of rubber are man-made materials so these characteristic parameters cause the complicated computation to obtain the COF of the shoe-floor exactly compared to reality.

Slip and its related problems are not new but necessary and potential when the working frequency and the appearance of human, robots moving on the PV panel to clean more and more [15]–[17]. For instance, the experiments of Several German, Rapra Technology Limited and SATRA, the standards DIN 4832 – 100, DIN 51097, DIN 51130 and a series of other standards (depending on the choice of each country) are published respectively to provide conditions for bare feet or rubber-soled footwear to prevent slipping on inclined, wet or oil-induced surfaces. All experiments are based on one of three principles: maximum tilt angle (ramp test), energy loss due to friction (pendulum method) and pulling force value generated by friction force (tribometer). Despite the clarity of such principles, it is a typical flaw in these experiments is to rely on test perceptions of test participants or on their subjective observations.

The demand for a design for the purpose of measuring slip through a tensile load cell to obtain a slipping force with acceptable sensitivity and enabling control factors of working conditions is very necessary. This paper describes the principle, and design of a test apparatus used to evaluate the slip resistance between a synthetic rubber trackpad and photovoltaic surface. These experiments are carried out by varying the inclination of the surface with respect to various wet/dry working conditions. The shore hardness of the synthetic rubber varies from A30 to A40, and the slip is confirmed when the value displayed on the tensile loadcell changes significantly. The experimental data is automatically fed back to the computer and stored for the next step. After all, this research is also completed well with the result being the standard of maximum tilt angle when using synthetic rubber with Shore hardness A30 – A40 to move without slipping on the surface of PV panels and ensure labor safety. As the frequency of human and robot work in the photovoltaic or solar energy industry is increasing in the recent year [17]–[19], the results of this research are even more necessary and practical.

## II. SYSTEM DESCRIPTION

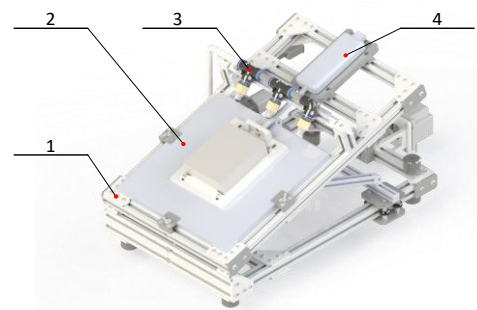

Figure 1.  3D design of the slip resistance test apparatus

The apparatus consists of four parts with separate functions to meet the above requirement: the chassis part (1), the photovoltaic panel (2), the water sprayer (3), and the tensile load cell (4). It works on the principle of transmission from the rotation of a direct current motor through a screw into the translation of casters. Then, the casters run along the slotted framing rails and the hinge enables the changing of the tilt angle of the upper surface of the part (1). Simultaneously, the angle, flow, and force sensors are sensed and fed to the computer for data logging.

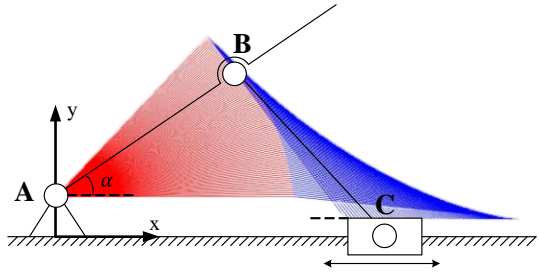

Figure 2. Kinematic diagram and operating simulation of the slip resistance test apparatus

The chassis part (1) consists of many slotted framing rails linked together through hard POM plastic sheets. The motor, the screw, the track, and the hinge are the driving devices, but the vibration of the motor acting on (1) will cause the vibration which degrades the measurement accuracy. Thus, adding shock-absorbing rubber soles is necessary. The water sprayer (3) has a flow sensor directly in front of it. "Working conditions" refer to water flow. Above the photovoltaic panel (2) there is a mass block, which is connected directly to (4) to detect when there is slippage occurs. The tensile load cell (4) consists of a tensile load cell with a sensitivity $0.1N$ and accuracy 0.2% [20], rigidly connected to a weight placed on either object of (2). With the force value displayed on the tensile load cell, the anti-slip standard will be calculated based on the expressions described in detail in the Basic principle of friction test method section below.

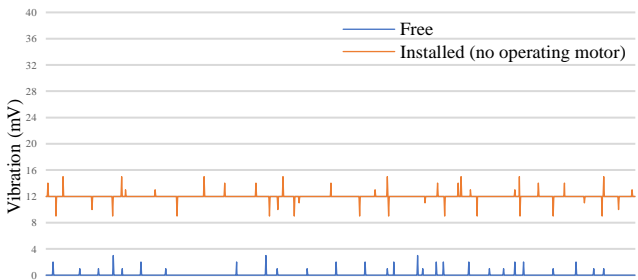

Figure 3. Test apparatus vibration in the free state

The test apparatus's vibration testing was staged based on feedback data from a Piezoelectric ceramic vibration sensor, which is placed on the bottom of the base and sends a signal to the computer. Although there are very low amplitudes of vibration in the free case (the sensor is placed alone in the air, not linked to any other thing) and in the installed case (no operating motor) (the sensor is linked to the chassis part (1) but the motor does not operate), they may be ignored due to their low frequency of appearance, so they can be regarded as signal noise of the sensor or singularities. Seeing that there is no vibration in the installed case (no operating motors larger than the free case), even though there is no other interaction between the two. Thus, this could explain that it was affected by the vibration of the ground, called a disturbance, and all test cases immediately after that influence are acceptable.

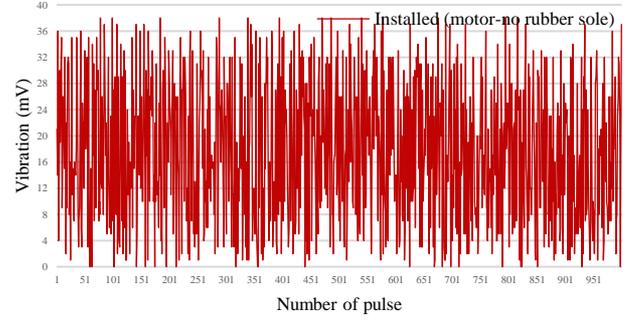

Figure 4. Test apparatus vibration when the rubber sole is not used

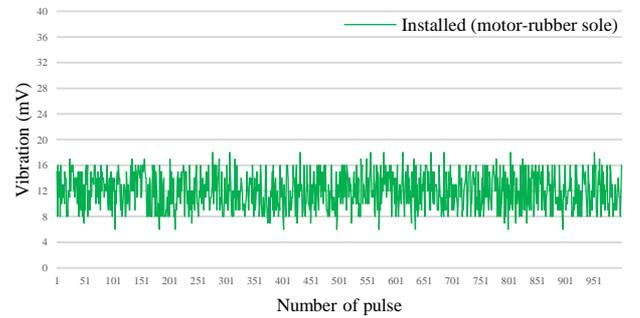

Figure 5. Test apparatus vibration when using a rubber sole

A proximity sensor that stops the caster C from exceeding the intended stroke regulates the safety of the entire apparatus. The usage of a stepper motor, vibration sensor, and angle sensor further guarantees precision with less than 5% inaccuracy. Equation (1) below depicts the relationship between the angle and the position of caster C.

$$AB \cos(\alpha) + \sqrt{BC^2 - [AB \sin(\alpha) + (y_A - y_C)]^2} = x_C \quad (1)$$

The parameters in the equation above mean:

- AB: length of link AB, shown in Fig. 2.
- BC: length of link BC, shown in Fig. 2.
- $\alpha$: tilt angle between the horizontal and upper surface.
- $y_A$: vertical coordinate of joint A.
- $y_C$: vertical coordinate of caster C.
- $x_C$: horizontal coordinate of caster C.

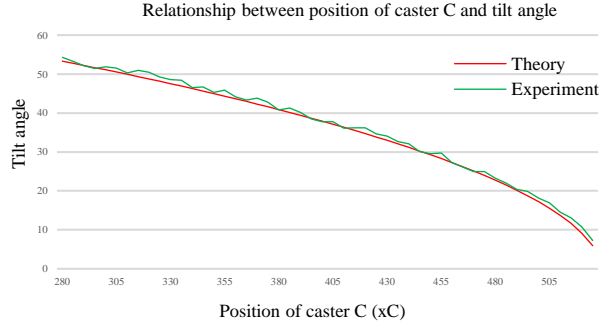

Figure 6. Theoretical and actual caster C's position data

## III. BASIC PRINCIPLE OF FRICTION TEST METHOD

These experiments are built on the theory that slip occurs when the magnitude of gravity $(\vec{P})$ is greater than the static friction $(\vec{F}_{fr})$ value, in the case of dry or wet environments, these values are also different.

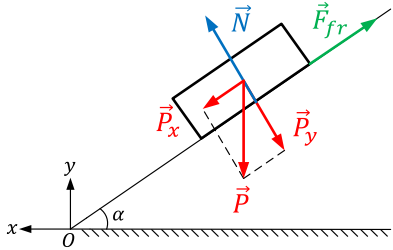

Figure 7. Force analysis in the absence of water force

$$F_{fr} \geq P \sin(\alpha) \qquad (2)$$

In the case of an additional traffic factor, there is a slight change. The force generated by the water $(\vec{F}_w)$ is directly proportional to the value of the flow rate and inversely proportional to the tilt angle.

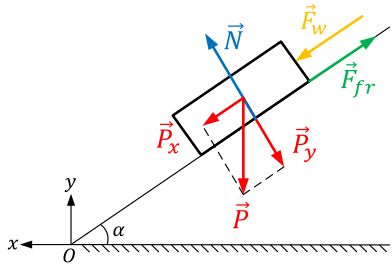

Figure 8. Force analysis in the presence of water force

$$F_{fr} \geq P \sin(\alpha) + F_w \qquad (3)$$

Thus, it can be seen that if no slip occurs, within a period of several seconds the value on the tensile loadcell does not change much and vice versa.

## IV. EXPERIMENTS AND RESULTS

Experiments perform for two pairs of subjects to find the maximum tilt angle that responds to the slip resistance of variable Shore A hardness synthetic rubber materials, these are Shore A30 – A35 with PV surface and Shore A35 – A40 PV surface. This rubber with a thickness of 10mm is placed under the same heavy block with a mass of 2kg.

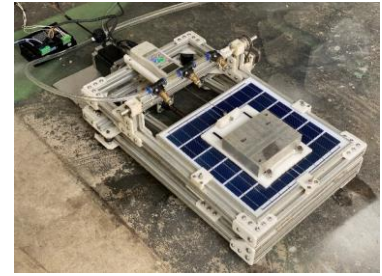

Figure 9. Actual image of experimental test apparatus

A total of 10 experiments were included when changing four values related to humid conditions. The angle changes with each screw step (5mm), stop for 10 seconds to observe the value on the tensile loadcell, if there is no change around the previously displayed value, no-slip has occurred but only the water force. Until there is a large change in the displayed force value, it means that the object is slipping. Then the experiments are adjusted to screw step at a smaller resolution (0.5mm) to get the most accurate angle value possible.

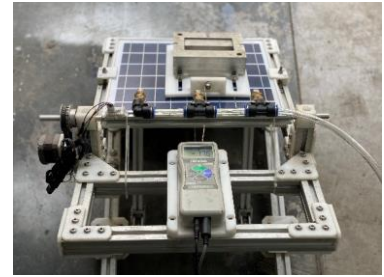

Figure 10. Actual image of test apparatus, slip detection at 29.3°

TABLE I. CONDITIONS AND RESULTS OF MAXIMUM TILT ANGLE WHEN TESTING THE RELATIONSHIP BETWEEN SHORE HARDNESS AND FLOW

| Flow rate | Shore A30 – A35 | Shore A35 – A40 |
|---|---|---|
| Dry | 29.3° | 25.5° |
| 0 l/min | 26.5° | 22.4° |
| 1 l/min | 24.7° | 20.5° |
| 2 l/min | 22.4° | 18.3° |
| 3 l/min | 21.2° | 17.0° |

TABLE I shows the maximum tilt angle results to avoid slip with different flow rates. As the above results, in the same value of shore A, if the flow rate is higher, the tilt angle will be lower. The relationship between flow rate and maximum tilt angle doesn't seem to be linear but like a logarithmic function and towards saturation. That needs more experiments to confirm.

A few minor error-related problems may occur due to the man-made nature of the material, mounting operation, measurement, and other environmental conditions (e.g. humidity, wind, dust, etc.). However, the above result is also supported in part by other related research and demonstrates that the relationship between Shore hardness and grip is inversely proportional.

TABLE II. MAXIMUM TILT ANGLE IS SUITABLE FOR SHORE HARDNESS TO ENSURE SAFETY

| Working condition | Shore hardness | Recommended tilt angle for safety |
|---|---|---|
| Dry | Shore A30 – A35 | < 29° |
| | Shore A35 – A40 | < 25° |
| 0 ÷ 3 l/min | Shore A30 – A35 | < 21° |
| | Shore A35 – A40 | < 17° |

After the experiment, the results are shown in TABLE II. These give some recommendations for rubber Shore A30 – A35, A35 – A40 when moving on the PV surfaces to ensure safety. With the different working conditions, Shore A30 – A35 should work with a tilt angle below 21° (wet) or 29° (dry) and below 25° (wet) or 17° (dry) for Shore A35 – A40 to avoid slipping.

## V. CONCLUSION

The article introduces a system that can detect and measure the sliding force of a pair of materials under various working conditions, from dry to varying humidity levels. Additionally, it prevents subjective influence from people, assuring objectivity. The study has been effective in providing the necessary maximum tilt angle that may be modified to prevent slip and fall incidents with neoprene materials and PV panels. From this, more studies may be done to determine the coefficient of friction for synthetic rubber at each Shore degree, under each operating circumstance, and while using safety standards to assure human safety.